\documentclass[showpacs,floatfix,twocolumn,prb,superscriptaddress]{revtex4-1}

\usepackage{graphicx,color}

\begin{document}

\title{Antimony arsenide: Chemical ordering in the compound SbAs}

\author{Daniel P. Shoemaker}
\affiliation{Materials Science Division, Argonne National Laboratory,
Argonne, IL 60439, USA}

\author{Thomas C. Chasapis}
\affiliation{Department of Chemistry, Northwestern University, Evanston, Illinois 60208, United States}

\author{Dat Do}
\affiliation{Department of Physics and Astronomy, 
Michigan State University, East Lansing, Michigan 48824, USA }

\author{Melanie C. Francisco}
\author{Duck Young Chung}
\affiliation{Materials Science Division, Argonne National Laboratory,
Argonne, IL 60439, USA}

\author{S. D. Mahanti}
\affiliation{Department of Physics and Astronomy, 
Michigan State University, East Lansing, Michigan 48824, USA }

\author{Anna Llobet}
\affiliation{Lujan Neutron Scattering Center, Los Alamos National Laboratory,
Los Alamos, NM 87545, USA}

\author{Mercouri G. Kanatzidis}\email{m-kanatzidis@northwestern.edu}
\affiliation{Materials Science Division, Argonne National Laboratory,
Argonne, IL 60439, USA}


\begin{abstract}
The semimetallic Group V elements display a wealth of correlated electron
phenomena due to a small indirect band overlap that leads to relatively
small, but equal, numbers of holes and electrons at the Fermi energy with
high mobility. Their electronic bonding characteristics produce a unique
crystal structure, the rhombohedral A7 structure, which accommodates lone
pairs on each site. Here we show that the A7 structure can display chemical
ordering of Sb and As, which were previously thought to mix randomly. Our
structural characterization of the compound SbAs is performed by single-crystal
and high-resolution synchrotron x-ray diffraction, and neutron and x-ray pair distribution
function analysis. All least-squares refinements indicate ordering
of Sb and As, resulting in a GeTe-type structure without inversion symmetry.
High-temperature diffraction studies reveal an ordering transition around 550 K.
Transport and infrared reflectivity measurements, along with
first-principles calculations, confirm that SbAs is a semimetal, albeit
with a direct band separation larger than that of Sb or As.
Because even subtle substitutions in the semimetals, notably Bi$_{1-x}$Sb$_x$,
can open semiconducting energy gaps, a further investigation of the interplay
between chemical ordering and electronic structure on the A7 lattice is warranted.
\end{abstract}



\maketitle

\section{Introduction} 

\begin{figure}
\centering\includegraphics[width=0.85\columnwidth]{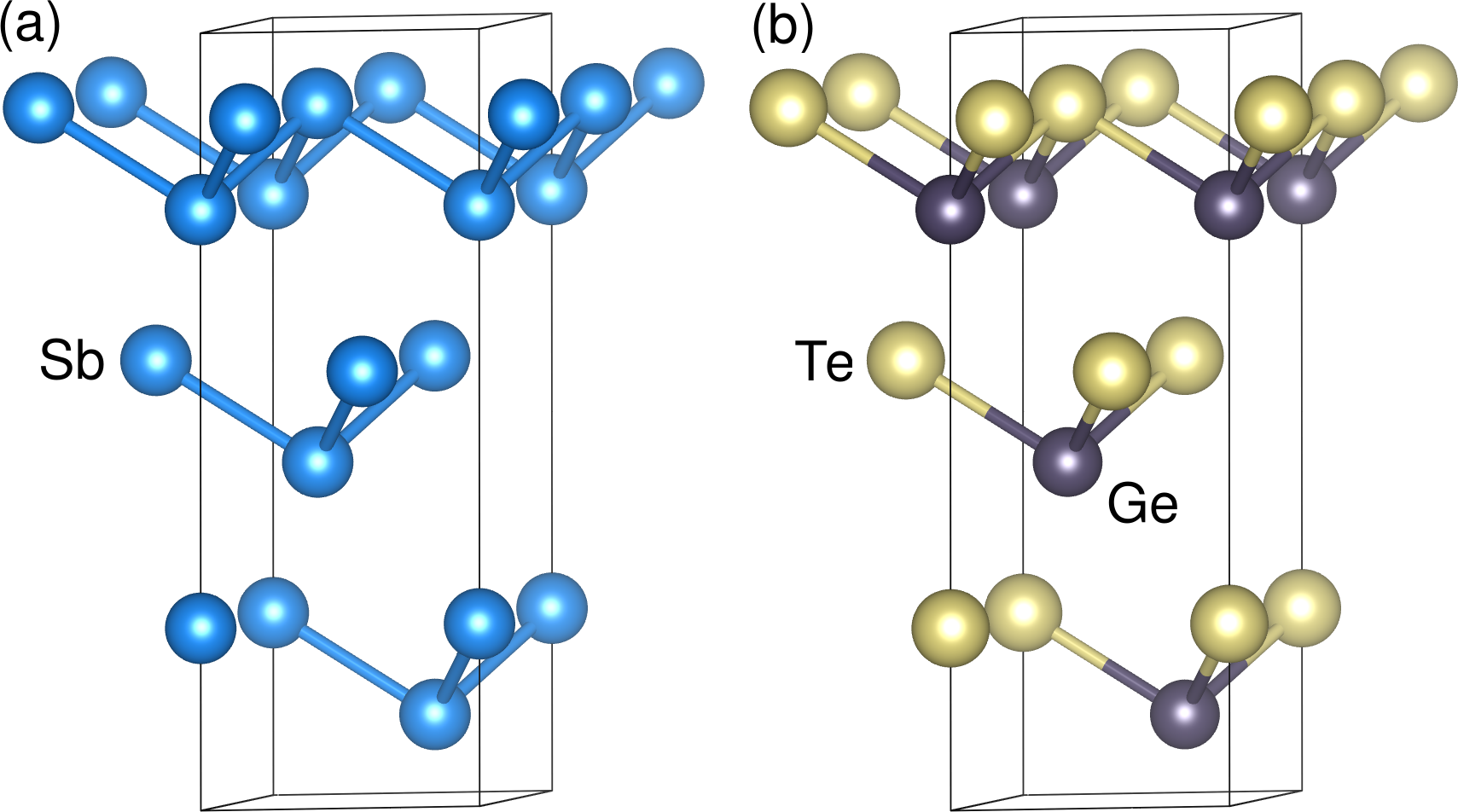} \\
\caption{(Color online) Unit cell of the A7 structure (a) of pure As, Sb,
and Bi (and black P at high pressure) with $R\overline{3}m$ symmetry. Ordering
the two $3a$ sites on this lattice leads to the GeTe structure in (b), lowering symmetry to
$R3m$, while retaining the same two $3a$ sites.
}
\label{fig:unitcells}
\end{figure}

The Group V elements As, Sb, and Bi have a structure all their
own, unshared by any other elements. This ``A7'' crystal structure 
in Figure \ref{fig:unitcells}(a) is 
a consequence of the unique electronic structure
of these elements, where $s$-$p$ hybridization leads to formation of a lone
pair. \cite{seo_what_1999,gonze_first-principles_1990}
As a result, the A7 structure has rhombohedral crystal symmetry 
and lies in space group $R\overline{3}m$. It is only a small distortion
removed from simple cubic symmetry, which can be experimentally
accessed under applied
pressure. \cite{da_silva_first_1997,degtyareva_high-pressure_2004,beister_rhombohedral_1990}
The band structure that drives formation of the A7 phase also
causes these elements to be the prototypical semimetals, with a small
offset band overlap, small number of carriers compared to typical
metals (10$^{-5}$ as many), high mobility, and nearly equal
concentrations of electron and hole carriers. \cite{saunders_semimetals_1973}

Their unique electronic properties have made the semimetals a
fascinating arena in condensed matter physics, permitting
initial measurements of the quantum mechanical oscillatory Shubnikov-de Haas and
de Haas van Alphen effects, Seebeck's discovery of the 
thermoelectric effect, and Hall's measurements of spin-dependent
transport.\cite{issi_low_1979}
Because the band overlap in semimetals is so delicate, they are 
tunable by doping for thermoelectric applications\cite{tritt_advances_2000}
and in topological insulators.\cite{hasan_colloquium_2010}

Alloying the Group V elements themselves can produce unexpected results.
The Bi$_{1-x}$Sb$_x$ solid solution is semimetallic on both ends
but becomes semiconducting for $0.07 < x < 0.22$ by
opening a gap at the $L$ point and removing
overlap at the $T$ point of the Brillouin zone.\cite{jain_temperature_1959,ibrahim_thermoelectric_1985}
Bi and As are not chemically miscible.\cite{predel_as-bi}
Previous work on Sb$_{1-x}$As$_x$ has found that these elements
are miscible across the full composition range.
\cite{saunders_electrical_1965,ohyama_electrical_1966}
Ohyama studied the thermal conductivity of Sb$_{1-x}$As$_x$ 
and speculated that an anomaly around $x$ = 0.5 might arise 
from an ordered compound but lacked any structural characterization 
to substantiate this claim.\cite{ohyama_thermal_1967}

While no pure elements outside Group V can form the A7 structure,
the isoelectronic IV-VI compound GeTe is closely related. It is simply an ordered 
arrangement of the same structure, shown in Figure \ref{fig:unitcells}(b), with
space group $R3m$ in contrast to $R\overline{3}m$ for A7. 
GeTe might make a promising ferroelectric were it not for large concentrations
of free carriers arising from Ge vacancies and a Fermi energy lying
within the valence band.\cite{steigmeier_soft_1970,bahl_amorphous_1970,rabe_structural_1987}
 Again, the GeTe structure is a consequence
of $s$-$p$ hybridization and lone pairs on both Ge and Te.\cite{fons_phase_2010}
No other pure compounds are known
to crystallize in the GeTe structure, as nearby IV-VI compositions form the GeS
or rocksalt structures, which are only small displacements
away.\cite{waghmare_first-principles_2003}

Distinguishing between GeTe-type chemical ordering ($R3m$) versus an 
A7 solid solution ($R\overline{3}m$) is very difficult---no new Bragg peaks
arise upon lowering symmetry, so careful analysis 
of high-quality scattering data is required. For that reason, and to make clear the
case for chemical ordering, we investigated SbAs using single crystal diffraction, high-resolution
synchrotron powder diffraction, and pair distribution function (PDF) refinements. 
All find preference for 80/20 ordering of Sb and As.
The PDFs do not show any short As-As distances which would
signal the onset of nanoscale
phase separation that is invisible to Bragg 
diffraction.\cite{petkov_high_1999,peterson_local_2001,bera_soluble_2010,shoemaker_total_2010}
High-temperature Bragg diffraction finds the atomic
mixing to become disordered above 300$^\circ$C.

Transport and reflectivity measurements confirm that SbAs is a semimetal
with band overlap, in agreement with our first-principles calculations.
While the Seebeck coefficient and plasma frequency of SbAs lie between
those of Sb and As, the optical dielectric constant is found to be 
outside the end members, implying a widening of the direct band
gap below the Fermi energy, indicating complex band structure
changes. In the case of BiSb, this 
band shifting leads to opening of a semiconducting energy gap.
None is found so far in SbAs, but the effects of doping and
annealing remain a topic of further investigation.

\section{Methods}

SbAs single crystals were prepared from elemental Sb (99.99\%) and As (99.999\%).
The powders were loaded into 9 mm-diameter quartz tubes and sealed under
vacuum. Tubes were heated at 10$^\circ$C/min to 800$^\circ$C, at which point the 
samples were molten. Samples were held at this temperature
for 30 min and periodically flipped to homogenize, then water
quenched to avoid an incongruent melting transition.
\cite{mansuri_equilibrium_1928}. Still sealed under vacuum, tubes were placed into a furnace
preheated to 630$^\circ$C to anneal for 60 h, then cooled
to room temperature at 10$^\circ$C/min. Under these conditions
SbAs crystallized into shiny, mirror-like crystals about 1 mm per side,
typically with triangular facets and
easily cleaved into plates along $\{001\}$ planes.

Single crystal X-ray diffraction data were collected up to $\theta$ = 35.28$^\circ$
on a STOE 2T image plate
diffractometer equipped with Mo-$K \alpha$ radiation ($\lambda$ = 0.71073 \AA) at 
room temperature. Data
reduction and integration absorption correction were performed using X-Area software
provided by STOE. The crystal structure was solved using direct methods and refined by a least-
squares refinement using the SHELXTL suite of programs.\cite{sheldrick_short_2007}
All atomic displacement parameters
were refined anisotropically. A twin law (-1 -1 0 0 1 0 0 0 -1) was applied and refined to
50.2\%. The final composition refined to Sb$_{0.94}$As$_{1.06}$. Crystallographic
parameters are given in Table \ref{tab:stoe}.

High-resolution synchrotron X-ray powder diffraction was performed at 
beamline 11-BM of the Advanced Photon Source (APS), using 30 keV x-rays
($\lambda = 0.413284$ \AA) and crystals
ground and sieved to 45 $\mu$m. High-temperature diffraction was
performed at beamline 1-BM using 20 keV x-rays ($\lambda = 0.6128$ \AA) 
and samples sealed under vacuum in quartz capillaries.
Time-of-flight neutron powder diffraction data were collected at the NPDF
instrument at Los Alamos National Laboratory.
Rietveld refinement was performed using the EXPGUI frontend\cite{toby_expgui_2001} for 
\textsc{GSAS}.\cite{larson_general_2000}. Unit cells and Fourier maps
are plotted using VESTA.\cite{momma_vesta_2008}

High-momentum-transfer total scattering data were collected
at APS beamline 11-ID-B (90 keV, $\lambda = 0.13702$ \AA), and the aforementioned NPDF
instrument (neutron time-of-flight).
Extraction of the PDF was performed using PDFGetX2\cite{qiu_pdfgetx2_2004} 
and $Q_{max}$ = 25\,\AA$^{-1}$
for x-ray data, and PDFGetN \cite{peterson_pdfgetn_2000} with $Q_{max}$ = 35\,\AA$^{-1}$
for neutron data. Least-squares fits to the PDF were conducted with PDFgui.\cite{farrow_pdffit2_2007}

Resistivity measurements were performed in 4-point geometry
using a Quantum Design PPMS.
The Seebeck coefficient (thermopower) of a polycrystalline SbAs ingot was
measured under helium atmosphere using an ULVAC-RIKO ZEM-3 system.
Samples used for infrared (IR) reflectivity were annealed at 630$^\circ$C 
for 30 h and cooled to room temperature in 60 h, resulting in 
large ingots.
A flat surface of the sample was alumina polished and washed with 
ethanol. The reflectivity spectrum was 
recorded as a function of wavenumber, in nearly normal incidence, 
in the spectral range 100-4000 cm$^{-1}$ with a Nicolet 6700 FTIR 
spectrometer equipped with a Spectra-Tech spectral reflectometer. 

Electronic band structures and densities of states (DOS) were
calculated for the hexagonal unit cell of 
SbAs using density functional theory (DFT). 
We used a perfectly ordered 
arrangement of SbAs but allowed for relaxation. All calculations 
used the projector-augmented wave  method \cite{bloch94, kresse99} 
and the generalized gradient approximation to exchange correlation, 
developed by Perdew-Burke-Ernzerhof, \cite{pbe} as implemented in the 
VASP code.\cite{vasp1,vasp2,vasp3} A planewave energy cutoff of 400 eV was 
used and convergence was assumed when the energy difference between 
subsequent self-consistent cycles was less than $10^{-4}$~eV. Self-consistent 
calculations were done using 12$\times$12$\times$6 Monkhorst-Pack 
\textbf{k}-point sampling\cite{monkhorst76} and the DOS is obtained by using a 
finer \textbf{k}-mesh of 18$\times$18$\times$9. Scalar relativistic effects 
and spin-orbit interactions were included. 
Thermopower $S$ calculations using the Boltzmann transport 
equation and constant relaxation time were performed with the 
BoltzTrap package written by Madsen and Singh.\cite{madsen_boltztrap_2006}

\section{Results and Discussion}

\subsection{Structural refinement}

\begin{table}
\caption{\label{tab:stoe} 
Structural parameters obtained from room-temperature
single-crystal refinement (full-matrix least-squares on $F^2$) of SbAs.
$R_1 = \Sigma ||F_o|-|F_c||/\Sigma|F_o|$,
$wR_2 = \{ \Sigma[ w(|F_o|^2-|F_c|^2)^2]/\Sigma[w(|F_o|^4)]\}^{1/2}$
}
\centering
\begin{tabular}{ll}
\hline
Formula								& Sb$_{0.94}$As$_{1.06}$\\
Formula Weight						& 193.86 g/mol\\
Crystal system						& Trigonal\\
Space group							& $R3m$\\
$a=b$									& 4.0655(7) \AA \\
$c$									& 10.889(3) \AA \\
$V$,$Z$ 								& 155.87(5) \AA$^3$, 3 \\
$\rho$ 								& 6.196 g/cm$^3$ \\
Absorption coefficient			& 28.748 mm$^{-1}$ \\
$F$(000) 							& 249 \\ 
$\theta_{max}$ 					& 34.62$^\circ$ \\
Reflections collected, unique	& 762, 197 \\
Unique reflections 				& 197 \\ 
$R_{int}$							& 0.0203 \\
Number of parameters 			& 11 \\
Goodness-of-fit on $F^2$		& 1.311 \\
Final $R$ indices [$I > 2\theta$($I$)] & 0.0117 $R_1$, 0.0287 $wR_2$\\
\hline
\end{tabular}
~\\
\end{table}

\begin{table*}
\caption{\label{tab:atoms} 
Atomic parameters obtained from single-crystal and synchrotron powder refinement of SbAs.
Atomic displacement parameters $U_{ij}$ are given in units of \AA$^2$.
For powder and PDF refinements, the total occupancy of each site was constrained to be 1,
but the antisite fraction was not constrained to be identical for both sites.
}
\centering
\begin{tabular}{p{1.7cm}p{1.7cm}p{1.7cm}p{1.7cm}p{1.7cm}p{1.7cm}p{1.7cm}p{1.7cm}p{1.7cm}}
\hline\hline
Atom   & $x$ & $y$ &        $z$ & Occupancy & $U_{11}=U_{22}$ &  $U_{33}$ &  $U_{12}$ & $U_{13}=U_{23}$ \\
\hline 
\multicolumn{9}{c}{single-crystal Mo-$K\alpha$ refinement} \\
\hline
Sb(1)  &   0 &   0 &          0 &   0.74(7) &        0.015(1) &  0.017(1) &  0.008(1) &               0 \\
As(1)  &   0 &   0 &          0 &   0.26(7) &        0.015(1) &  0.017(1) &  0.008(1) &               0 \\
As(2)  &   0 &   0 &  0.4623(1) &   0.80(9) &        0.016(1) &  0.017(1) &  0.008(1) &               0 \\
Sb(2)  &   0 &   0 &  0.4623(1) &   0.20(9) &        0.016(1) &  0.017(1) &  0.008(1) &               0 \\
\hline
\multicolumn{9}{c}{powder 11-BM synchrotron refinement} \\
\hline
Sb(1)  &   0 &   0 &          0 &  0.799(4) &       0.0121(3) & 0.0163(8) & 0.0061(1) &               0 \\
As(1)  &   0 &   0 &          0 &  0.201(4) &       0.0121(3) & 0.0163(8) & 0.0061(1) &               0 \\
As(2)  &   0 &   0 & 0.46130(6) &  0.783(5) &       0.0158(4) &  0.019(1) & 0.0078(2) &               0 \\
Sb(2)  &   0 &   0 & 0.46130(6) &  0.217(5) &       0.0158(4) &  0.019(1) & 0.0078(2) &               0 \\
\hline
\multicolumn{9}{c}{X-ray PDF least-squares refinement} \\
\hline
Sb(1)  &   0 &   0 &          0 &      0.84 &          0.0138 &    0.0158 &     0.007 &               0 \\
As(1)  &   0 &   0 &          0 &      0.16 &          0.0138 &    0.0158 &     0.007 &               0 \\
As(2)  &   0 &   0 &      0.462 &      0.89 &          0.0228 &    0.0197 &     0.011 &               0 \\
Sb(2)  &   0 &   0 &      0.462 &      0.11 &          0.0228 &    0.0197 &     0.011 &               0 \\
\hline \hline
\end{tabular}
~\\
\end{table*}

Single-crystal diffraction provides the most observed reflections per 
refined parameter and so is the preferred method to determine
whether SbAs forms an ordered compound  or solid solution.
The cell parameters and refinement details are
summarized in Table \ref{tab:stoe}, while the atomic parameters are given 
in Table \ref{tab:atoms}. It must be noted that lowering symmetry from the 
A7 solid solution space group $R\overline{3}m$
to GeTe-type $R3m$ does not result in the appearance of any new reflections. Only
the peak \emph{intensities} hold information about chemical ordering.
We found that the cell contains two sites with different scattering 
density. The refinement indicated that an approximate 80/20 occupation of each
site (see Table \ref{tab:atoms}) significantly improves the fit versus a pure 
solid solution: refinements
with even site mixing in the $R\overline{3}m$ space group
gave final $R$ indices of $R_1$ = 0.0270 and 
$wR_2$ = 0.0754 with a goodness-of-fit on $F^2$ of 1.527.
This implies that
SbAs has GeTe-type ordering with $\sim$20\% antisite disorder.
More complete ordering might be attainable by annealing below the
ordering temperature which we discuss subsequently.

\begin{figure}
\centering\includegraphics[width=0.85\columnwidth]{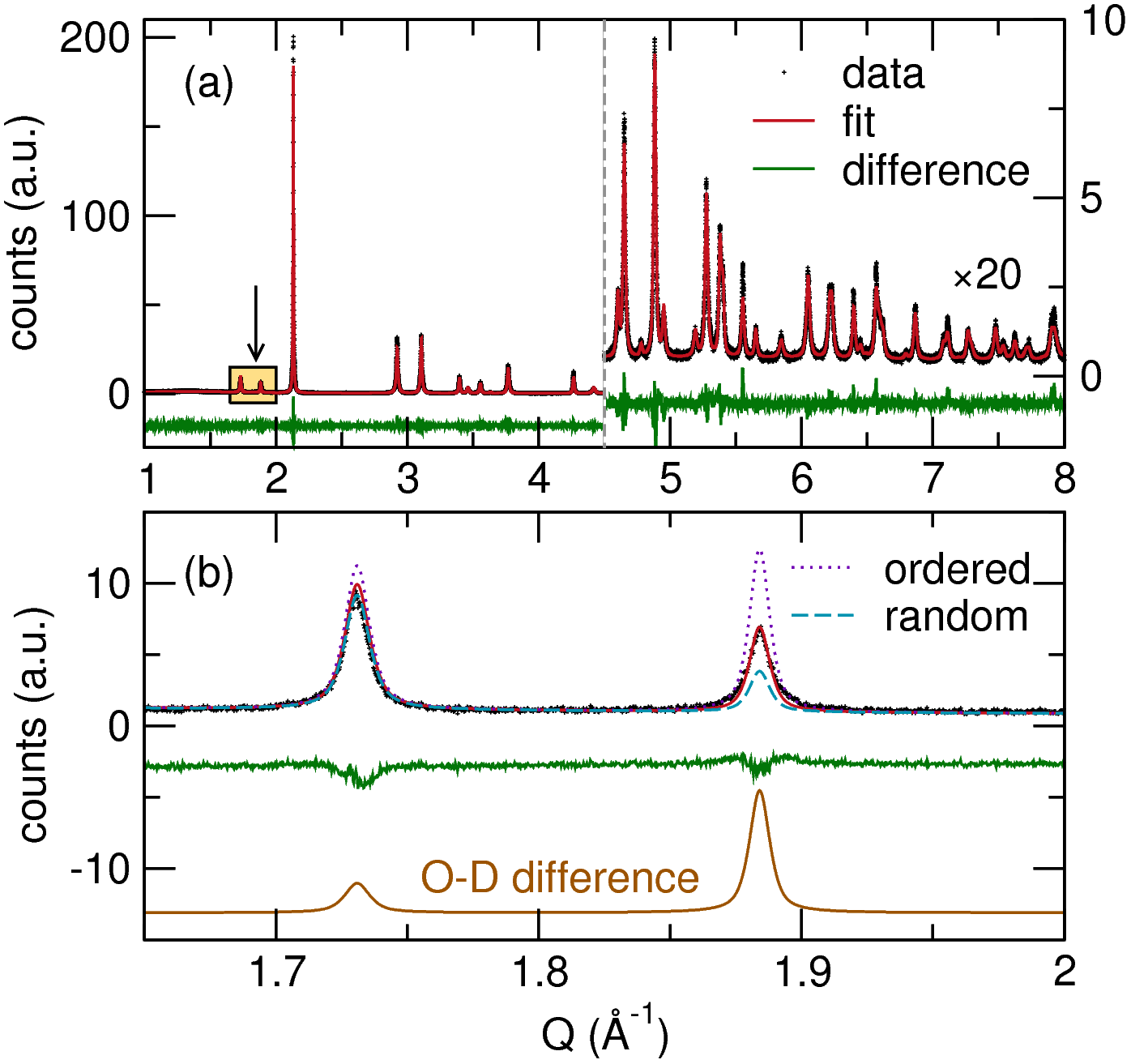} \\
\caption{(Color online) Rietveld refinement of SbAs using synchrotron X-ray radiation at beamline
11-BM of the APS. The full refinement range is given in (a), with the high-$Q$
region enlarged to show detail. Ordering between Sb/As produces a change
in intensity in the $\{101\}$ peak at $Q$= 1.88 \AA$^{-1}$, highlighted and arrowed. This area
is enlarged in (b), where the experimental and refined pattern is compared to
the model using fully-ordered (dotted) and random solid solution (dashed).
The difference between the two models (O-D difference)
is shown at the bottom of (b). The largest intensity variation in the pattern
lies on the $\{101\}$ peak.
}
\label{fig:rietveld-11bm}
\end{figure}

Single crystal refinement indicates that SbAs is
mostly ordered, but the two structures produce such similar diffraction
patterns that multiple probes should be used to confirm correct occupancy.
To that end, we performed synchrotron powder diffraction
at beamline 11-BM of the APS, which provides unparalleled resolution
and signal/noise ratio for powder samples.\cite{wang_dedicated_2008}
The 11-BM data is shown in 
Figure \ref{fig:rietveld-11bm}
and the Rietveld fit is excellent.
The 11-BM refinement comes to the same occupancies and $U_{ij}$
atomic displacement parameters as the single-crystal
refinement, within the margins of experimental error (Table \ref{tab:atoms}).

Powder diffraction patterns allow us to visualize
how much intensity is attributable to chemical ordering:
the difference
in the diffraction patterns for ordered and solid-solution SbAs
models 
(``order-disorder difference'')
is plotted at the bottom of each pane in Figure \ref{fig:rietveld-11bm}.
The difference was within the noise of conventional 
Cu-$K\alpha$ diffraction data. The region with the largest order-disorder
difference is highlighted in Figure \ref{fig:rietveld-11bm}(a) and 
magnified in Figure \ref{fig:rietveld-11bm}(b).
Here, the exceptional signal/noise ratio of 11-BM data provides 
distinction between the ordered (dotted) and disordered (dashed) models
in the $\{101\}$ peak at $Q$ = 1.88 \AA$^{-1}$. From this view
the increased experimental intensity versus the disordered model is clear, 
providing strong evidence for ordering. 

\begin{figure}
\centering\includegraphics[width=0.85\columnwidth]{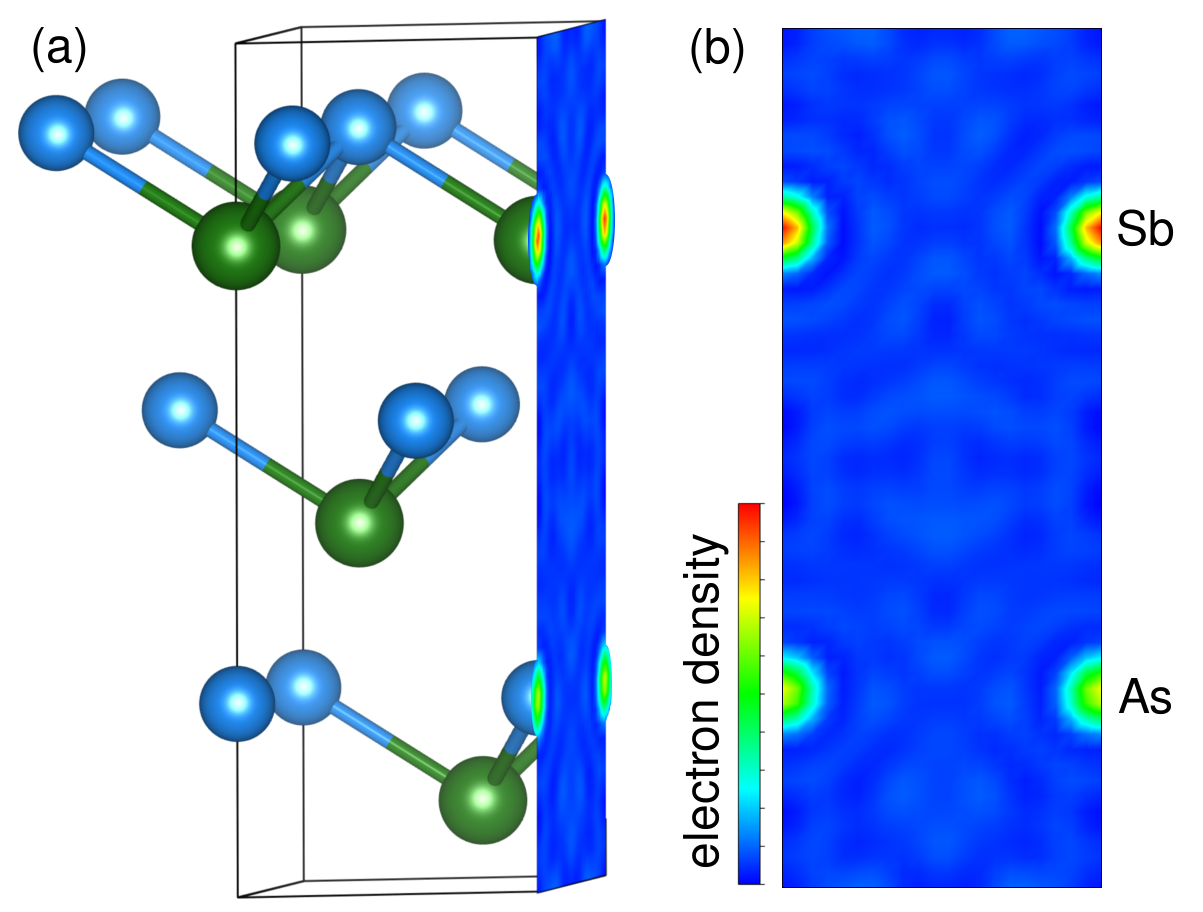} \\
\caption{(Color online) The refined structure (a) from 11-BM 
X-ray data is shown with a Fourier map of observed structure factors $F_{obs}$
on the $\{100\}$ face. 
The slice is viewed normal to the plane in (b), with  
stronger scattering density evident on Sb versus As sites. 
}
\label{fig:fourier}
\end{figure}


Since the diffraction pattern is the Fourier transform of the scattering
potential and the experimental structure factors $|F_{obs}|$ can be extracted 
directly from the 11-BM data, a Fourier
map can be constructed that contains the three-dimensional distribution
of scattering density in the cell.

 The $\{100\}$ slice of this map is 
attached to the side of the unit cell in Figure \ref{fig:fourier} to show the relationship
between the Sb and As sites and their location on the map. The $\{100\}$ Fourier map 
viewed normal to the plane in Figure \ref{fig:fourier}(b). Heavier
scattering is evident on the Sb site ($Z=51$ versus $Z=33$ for As). In a solid solution
these two sites would have equal scattering density.

\begin{figure}
\centering\includegraphics[width=0.85\columnwidth]{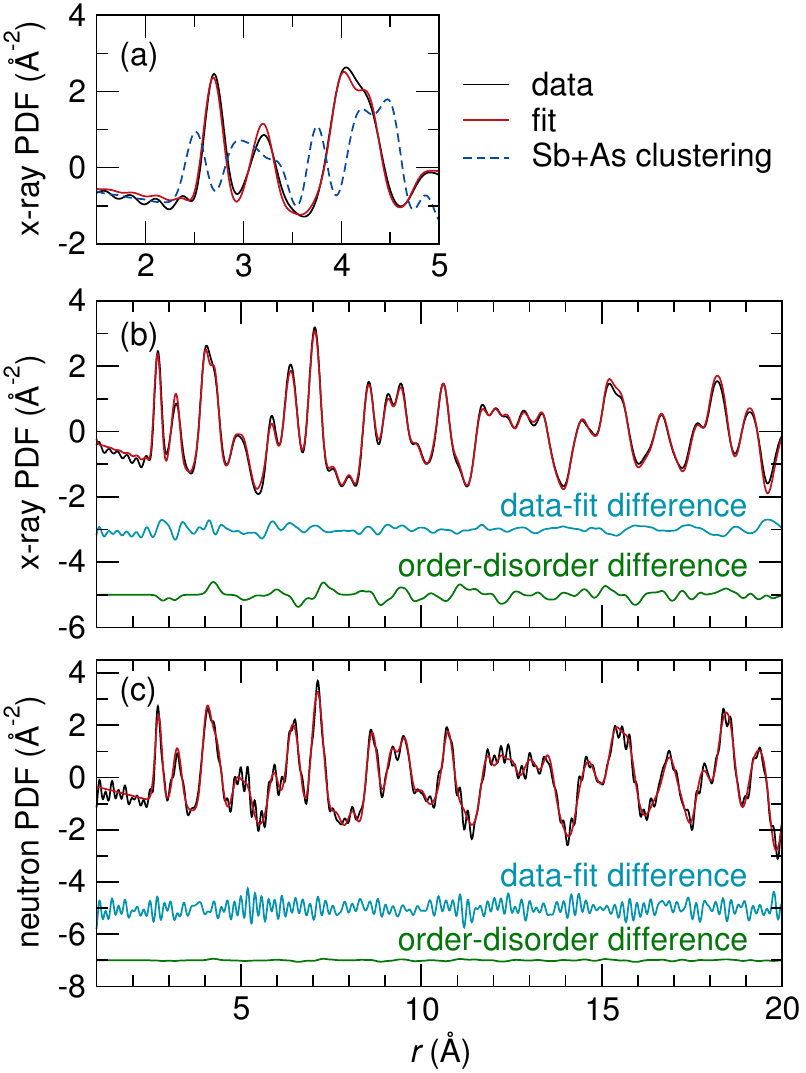} \\
\caption{(Color online) Least-squares refinements to the SbAs PDF collected at room
temperature using (a,b) x-rays and (c) neutrons. In both cases, fits are
excellent. The inset in (a) shows a two-phase mixture of Sb and As (dashed)
that would indicate clustering. This is not present in the data.
In the x-ray PDF,
the calculated difference between ordered and solid-solution SbAs is
larger than the noise. In the neutron case, due to similar scattering
lengths, this difference is well below the noise threshold.
}
\label{fig:pdfgui-fits}
\end{figure}

The presence of nanoscale heterogeneity in SbAs cannot be entirely
excluded by single crystal and powder refinements, so we turned
to PDF analysis.
Local-structure PDF studies of the supposed solid solutions
(In,Ga)As, Zn(Se,Te), and (Li,Na)AsSe$_2$ have shown that materials with single-phase
Bragg diffraction patterns can exhibit nanoscale clustering
of the end members, evidenced by split nearest-neighbor
bond distances in the PDF.\cite{petkov_high_1999,peterson_local_2001,bera_soluble_2010}
Could there be nanoscale clustering of Sb-rich and As-rich regions, as has been
proposed by Levin, et al?\cite{levin_effect_2003}
For extensive nanoclustering of Sb and As we would
see short bonds corresponding to As--As
(2.52 \AA), plus long Sb--Sb bonds (2.91 \AA).
The low-$r$ region of the experimental x-ray PDF in Figure 
\ref{fig:pdfgui-fits}(a) shows a fit to the single-phase model from Rietveld refinement,
plus a dashed line corresponding to the nanoclustered model.
None of the distinct As--As or Sb--Sb bonds are present, eliminating
the possibility of extensive nanoscale phase separation.

A unit cell model can be least-squares refined using the PDF, just as
was performed using single crystal and powder diffraction data.
The x-ray PDF gives the same refined occupancy and atomic parameters as
the single-crystal and powder refinements, providing a third
check of the Sb/As ordering. The fit is shown in Figure \ref{fig:pdfgui-fits}(b)
and refined values are given in Table \ref{tab:atoms}.
Neutron scattering does not provide sufficient contrast
between Sb and As (5.57 and 6.58 fm scattering cross-sections, respectively) 
to resolve site ordering in the PDF or Bragg peaks, but higher
$r$-space resolution in the neutron PDF (a result of higher usable $Q_{max}$)
further confirms the absence of nanoscale phase
separation by the good fit at low $r$ in Figure \ref{fig:pdfgui-fits}(c).


\begin{figure}
\centering\includegraphics[width=0.9\columnwidth]{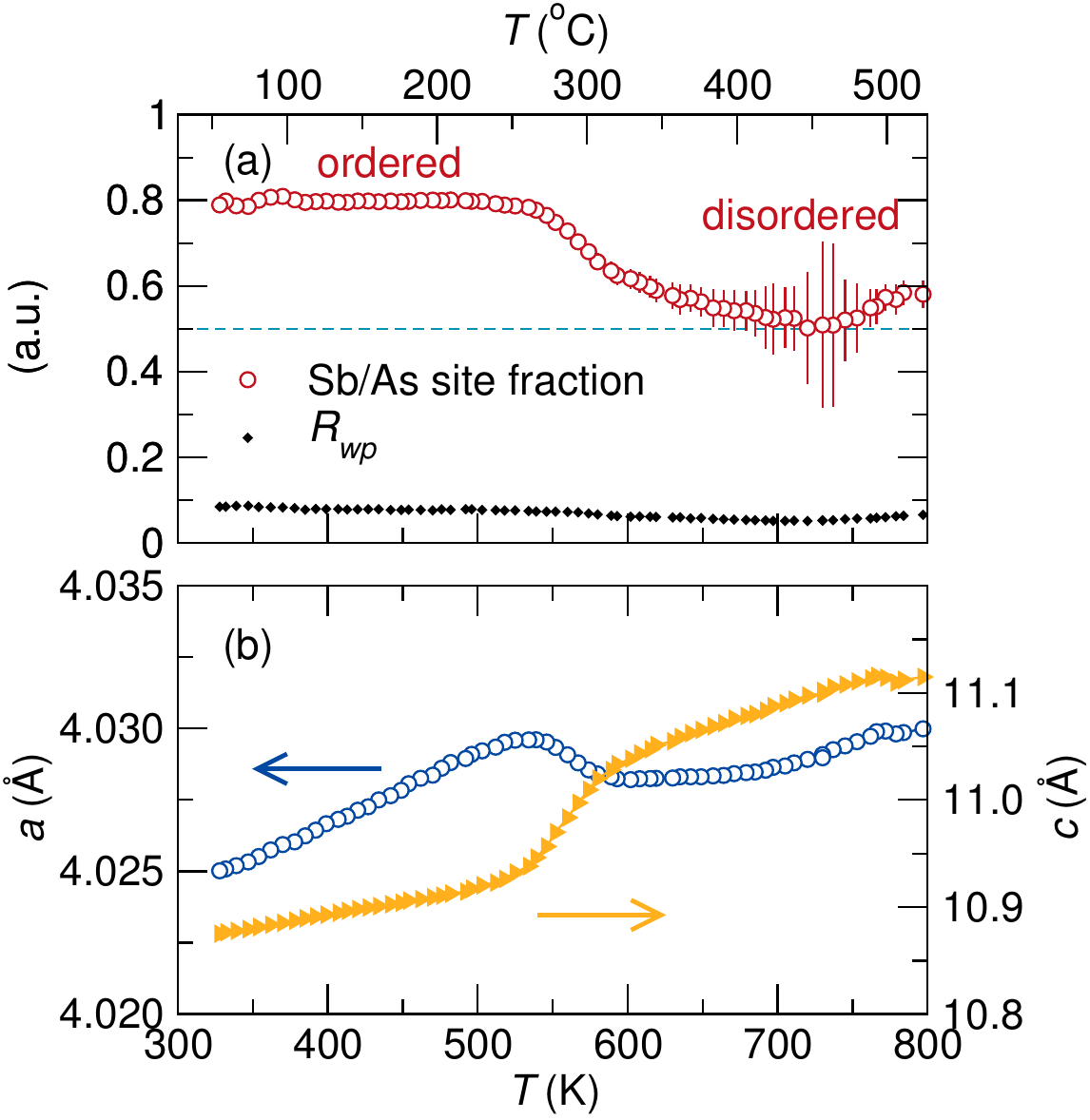} \\
\caption{(Color online) Sequential Rietveld refinements to high-temperature
synchrotron x-ray diffraction data (beamline 1-BM, APS)
show disordering of Sb/As around 550 K, evidenced
by the site occupancy in (a), while maintaining low goodness
of fit $R_{wp}$. In (b), the temperature dependence of the 
lattice parameters indicates a first-order
transition around 550 K.
}
\label{fig:1bm}
\end{figure}

High-temperature diffraction data collected upon heating a sample of
SbAs at the APS beamline 1-BM refine to the same 80/20 site ordering as 
other techniques, and results of sequential refinements upon 
heating are shown in Figure \ref{fig:1bm}. The site occupancy 
begins to deviate from the room-temperature  value around 
500 K. A value of 0.5 in Figure \ref{fig:1bm}(a) 
corresponds to even mixing (disorder) on both sites, and is
marked with a dashed line. Figure 
\ref{fig:1bm}(b) shows a rapid change in the lattice parameters
in the interval 500 K $< T <$ 600 K, suggesting a first-order transition
around 550 K. Annealing samples below this transition temperature
may result in more complete chemical ordering.

\subsection{Transport measurements}

\begin{figure}
\centering\includegraphics[width=0.85\columnwidth]{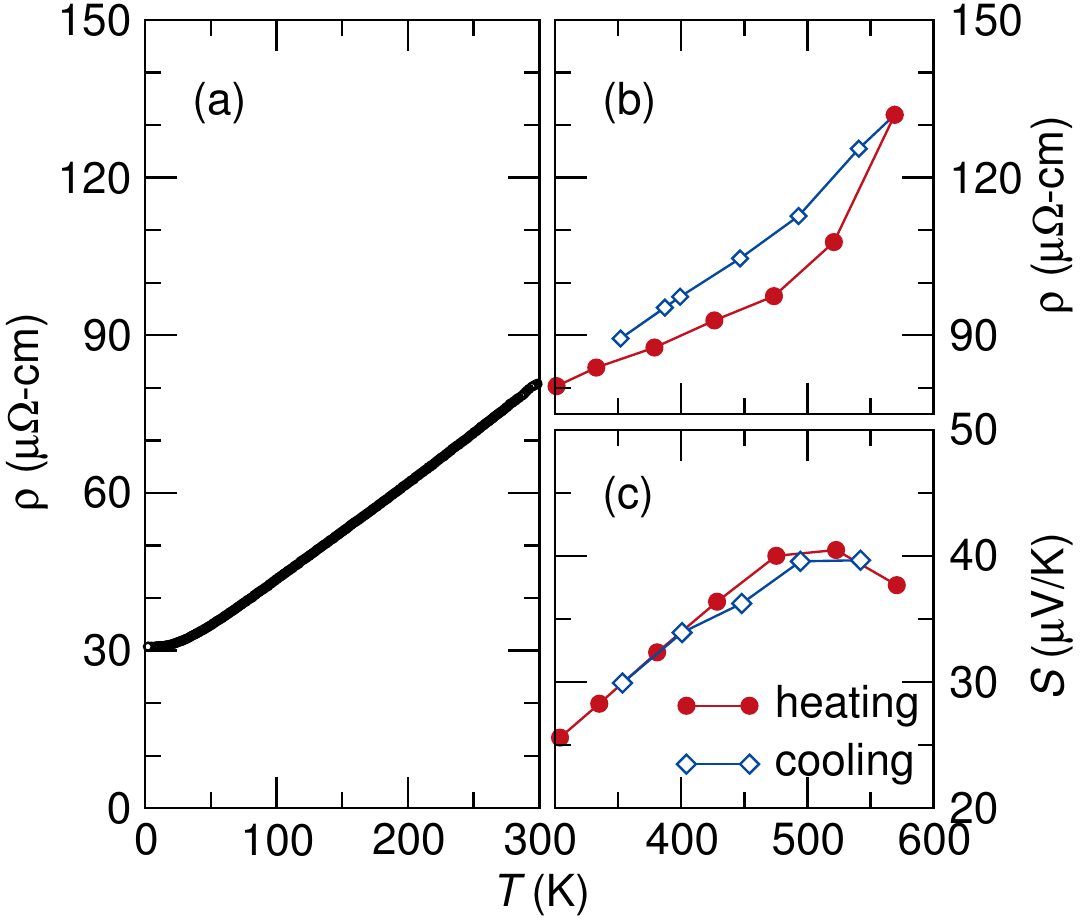} \\
\caption{(Color online) Resistivity (a) of SbAs shows impurity scattering at low
temperature and linear temperature dependence, typical of semimetals.
Resistivity of an ingot at higher temperatures (b) displays hysteresis, likely
due to the phase transition around 550 K.
Concurrent measurement of the Seebeck coefficient $S$ (c) shows $p$-type
conductivity with a broad maximum around 500 K. 
}
\label{fig:resistivity}
\end{figure}

Resistivity measurements of SbAs crystals in the $ab$ plane display
metallic behavior with $\rho$ = 80 $\mu\Omega$-cm at room temperature, 
shown in Figure \ref{fig:resistivity}(a). The large
residual resistivity at low $T$ is characteristic of impurity scattering.
There is no region where the temperature coefficient of resistivity is negative,
as would be expected for a 
semiconducting region with finite $E_g$. The behavior of SbAs can be
contrasted with the Bi$_{1-x}$Sb$_x$ system, where 7-22\% Sb substitution
leads to an opening of $E_g$ up to 0.014 eV.\cite{jain_temperature_1959,ibrahim_thermoelectric_1985}
Saunders qualitatively suggested that SbAs does not become
a semiconductor because the band overlap in As is greater than
that in Bi, so a larger perturbation from Sb addition would be
required to shift the $L$-point band enough to create a gap.\cite{saunders_electrical_1965}
Figure \ref{fig:resistivity}(b) shows a rise in the resistivity 
above 500 K, coincident with the order-disorder
transition observed by high-temperature diffraction in Figure \ref{fig:1bm}.

The measured thermopower of a polycrystalline
SbAs ingot is shown in Figure \ref{fig:resistivity}(c). The positive
($p$-type) behavior and magnitude are similar to that of pure Sb,
which reaches a maximum of 30-55 $\mu$V/K at 400 K, depending
on crystalline orientation.\cite{saunders_seebeck_1968}
The Seebeck coefficient of arsenic is about a factor of 4 smaller.\cite{saunders_seebeck_1965}
The decrease in $S$ at high temperatures indicates an increase
in $n$-type character, which could result from the effects
of different mobilities for electrons and holes or changes in band overlap due to 
lattice expansion. 
Given the intricacies of the band structure near $E_F$ in semimetals,
we turn to IR measurements and DFT calculations to understand
how chemical ordering in SbAs may affect the band structure.

\subsection{Infrared reflectivity}

\begin{figure}
\centering\includegraphics[width=0.85\columnwidth]{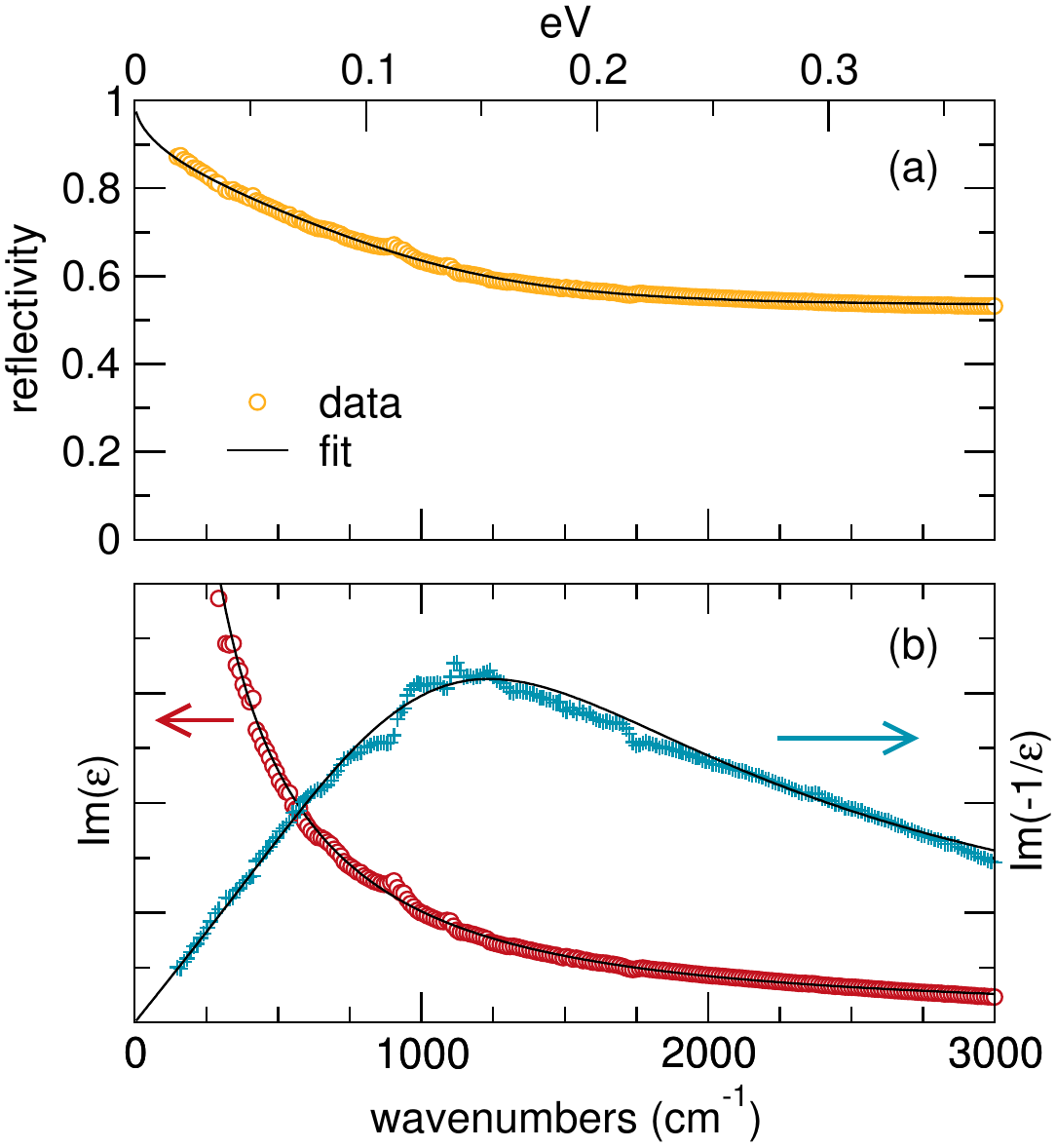} \\
\caption{(Color online) Experimental reflectivity spectrum (a) of SbAs 
and best-fit calculated reflectivity (solid line) using Equations \ref{eq:reflectivity}
and \ref{eq:drude}.
The Kramers-Kronig obtained Im($\varepsilon$) and the energy loss 
function Im($-1/\varepsilon$) are shown in (b). 
The Drude fit (Equation \ref{eq:drude}) to Im($-1/\varepsilon$) gives
the plasma frequency $\omega_p$ and complex dielectric function $\varepsilon(\omega)$.
}
\label{fig:ir1}
\end{figure}

The experimental reflectivity spectrum of SbAs is shown in Figure 
\ref{fig:ir1}. This spectrum is dominated by a structureless plasmon, yielding 
high reflectivity values in the low frequency range with a broad 
minimum around $\sim$1300 cm$^{-1}$, followed by almost constant reflectivity 
values in the high frequency regime. The spectrum was analyzed by 
the Kramers-Kronig method to obtain the real $\varepsilon_1$ and the 
imaginary $\varepsilon_2$ parts of the complex dielectric function $\varepsilon(\omega)$. 
The $\varepsilon_2$ spectrum Im($\varepsilon$) and the energy loss function 
\begin{equation}
-\mathrm{Im}\left( 1/\varepsilon \right) = \varepsilon_2 / \varepsilon_1^2 + \varepsilon_2^2
\end{equation}
are shown 
in Figure \ref{fig:ir1}. The Im($\varepsilon$) 
spectrum is increased in the low frequency range denoting the 
contribution of free carriers, while the peak in  Im($-1/\varepsilon$)  
indicates a plasma frequency of $\sim$1100 cm$^{-1}$. The latter represents 
the frequency of a longitudinal collective mode, when the entire 
carrier gas system is displaced relative to the fixed ions. 

The reflectivity $R(\omega)$ is expressed through the complex 
dielectric function as
\begin{equation}\label{eq:reflectivity}
R(\omega) = \left( \frac{ \sqrt{\varepsilon(\omega)}-1 }{ \sqrt{\varepsilon(\omega)}+1 } \right)^2
\end{equation}
We find that we can fit the measured reflectivity spectrum of 
Figure \ref{fig:ir1}(a) with a single-carrier double-damped Drude formula for 
the complex dielectric function \cite{gervais_optical_2002}
\begin{equation}\label{eq:drude}
\varepsilon(\omega) = \varepsilon_\infty \left( 1- \frac{ \omega_p^2 - i(\gamma_p - \gamma_0)\omega }{ \omega(\omega + i\gamma_0) } \right)
\end{equation}
where $\varepsilon_\infty$ is the optical dielectric 
constant associated with the bound electrons and $\omega_p$ 
is the plasma frequency:
\begin{equation}\label{eq:omegap}
\omega^2_p = \frac{Ne^2}{\varepsilon_0 \varepsilon_\infty m^*}
\end{equation}
where $N$ is the free carrier concentration and $m^*$ is the carrier 
effective mass. In the 
typical Drude expression for $\varepsilon(\omega)$, the free 
carrier damping factor $\gamma_p = 1/\tau$ is considered 
constant throughout the entire frequency range. In the case 
of Equation \ref{eq:drude}, the carrier relaxation time $\tau$ is taken to 
be frequency dependent, giving a frequency dependent 
damping factor. Here  $\gamma_p$ represents the linewidth of
the plasma response centred at $\omega = \omega_p$ and 
$\gamma_0$ represents the linewidth of the absorption at 
$\omega=0$.  The ratio $\gamma_p / \omega_p$  describes the 
motion of charge carriers: vibrational when the ratio is small, 
diffusive or incoherent when the ratio is large. Notice that 
Equation \ref{eq:drude} reduces to the simple 
Drude expression when  $\gamma_p = \gamma_0$.\cite{pessaud_optical_2000}
Fitting the data in Figure \ref{fig:ir1} gives the parameters in Table \ref{tab:ir}.
The single-carrier model fits our data well and also describes 
the IR reflectivity spectra of Bi$_{1-x}$Sb$_x$ alloys in the 
semimetallic composition range.\cite{stepanov_plasmon_2004,grabov_reflection_2001}
This, however, does not exclude a two-carrier system, i.e. electrons and holes,
because a two-carrier Drude expression with plasma frequency given by Equation
\ref{eq:twocarrier} is equivalent to that of a single carrier if the two 
carriers have the same relaxation time.
\begin{equation}\label{eq:twocarrier}
\omega^2_p = \frac{e^2}{\varepsilon_0 \varepsilon_\infty}\left(\frac{N_e}{m^*_e} + \frac{N_h}{m^*_h}\right)
\end{equation}

\begin{table}
\caption{\label{tab:ir} 
Drude parameters derived from room-temperature reflectivity data of SbAs
}
\centering
\begin{tabular}{p{1.8cm}p{1.8cm}p{1.8cm}p{1.8cm}}
\hline
$\omega_p$ (cm$^{-1}$)	&	$\varepsilon_\infty$	&	$\gamma_p$ (cm$^{-1}$)	&	$\gamma_0$ (cm$^{-1}$)	\\
1323							&	41.6						&	2044							&	1187 \\
\hline
\end{tabular}
~\\
\end{table}

For SbAs the plasma frequency $\omega_p$ = 1323 cm$^{-1}$ 
lies  between the two end members:  1000 cm$^{-1}$ for Sb,\cite{fox_optical_1974} 
and 2419 or 2017 cm$^{-1}$ for two different orientations of As.\cite{riccius_plasma_1971}
Using the values of Table \ref{tab:ir} and Equation \ref{eq:omegap}
we calculated $N/m^*$ = $8.14\times10^{20}$ cm$^{-3}$
for SbAs, which also lies between the respective values for Sb and As ($5.9\times10^{20}$ and
8.9$\times10^{21}$ cm$^{-3}$,
respectively).\cite{fox_optical_1974,riccius_plasma_1971,riccius_proc_1968}

Meanwhile, the optical dielectric constant $\varepsilon_\infty$ for SbAs
was found to be 41.6, which is less than the values for
Sb and As (80 and 50, respectively).\cite{harris_optical_1964,riccius_proc_1968}
The polarizability of the valence electrons relative to both parent
materials is decreased, which could be explained by the empirical Moss 
relation, $\varepsilon_\infty \propto E_g^{-1}$, which implies
that  $\varepsilon_\infty$ decreases as the direct band gap
$E_g$ increases. This relation holds
in Bi$_{1-x}$Sb$_x$ alloys across the band-opening composition region.
\cite{stepanov_plasmon_2004,grabov_dielectric_1990}
A similar direct gap opening opening at the $L$ point of SbAs, albeit 
small and below $E_F$, is found in our DFT electronic
structure calculations of SbAs.

Our IR reflectivity measurements confirm that SbAs displays
carrier concentrations typical of Sb and As, with band
overlap typical of a semimetal. However, the direct band
separation probed by $\varepsilon_\infty$ seems to have opened
wider than that of the end members. We conducted first-principles calculations
to visualize how chemical ordering affects the band structure.

\subsection{Electronic structure calculations}

\begin{figure}
\centering\includegraphics[width=0.95\columnwidth]{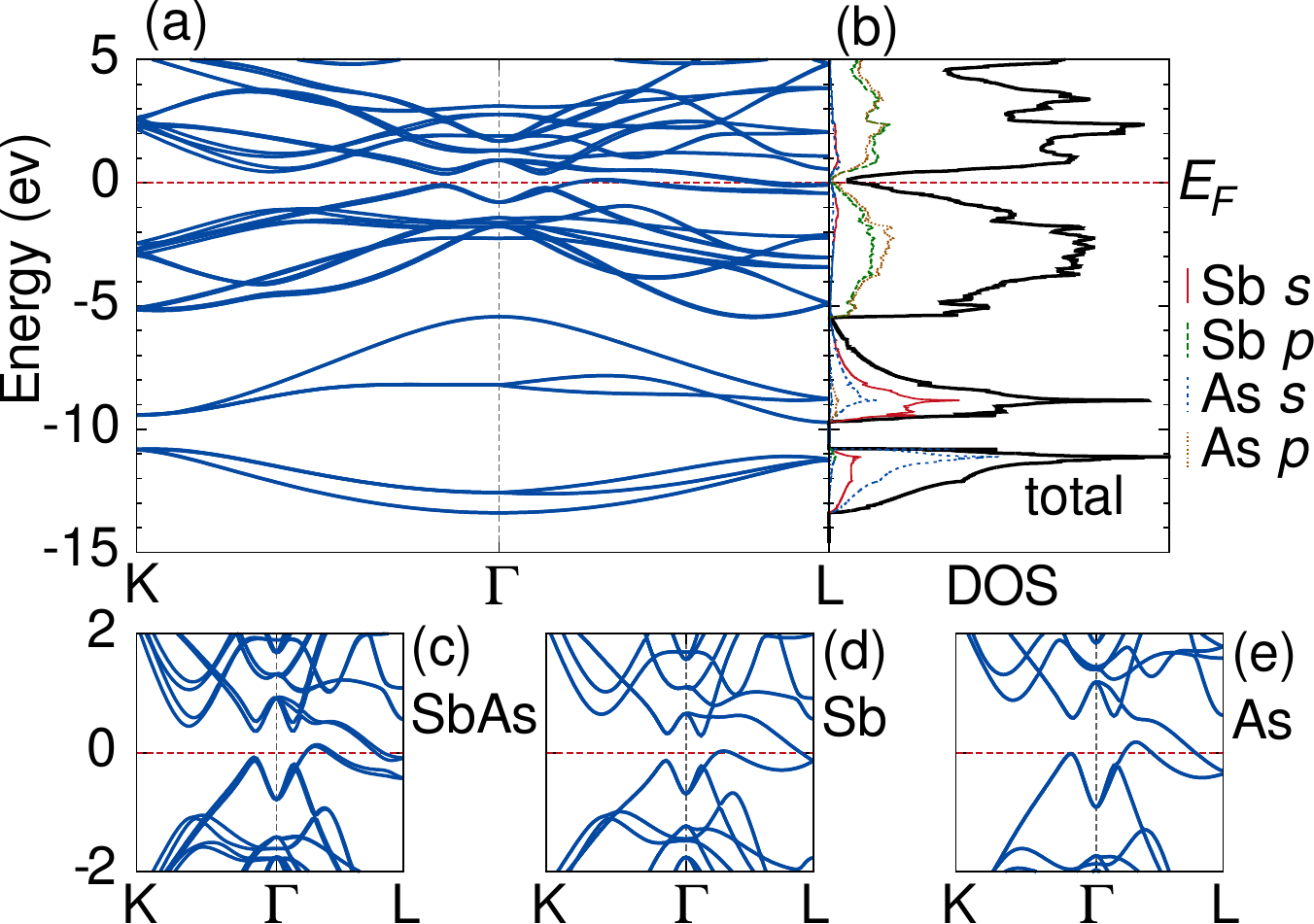} \\
\caption{(Color online) 
Electronic structure of SbAs (relativistic and spin orbit interaction 
are included): (a) Band structure with a pseudogap along the $\Gamma L$ 
direction; (b) projected density of states (DOS) showing a finite value
at $E_F$. Detailed band structures around $E_F$ are shown below for (c) SbAs, (d) pure Sb,
and (e) As.
}
\label{fig:bandstructure}
\end{figure}

The calculated band structure and DOS of SbAs are shown in Figure \ref{fig:bandstructure}(a,b). 
Our calculations clearly show that SbAs is a semimetal with a pseudogap, 
in agreement with resistivity and IR experiments. The detailed band structure along K$\Gamma$L 
near the pseudogap is given in the Figure \ref{fig:bandstructure}(c). The 
valence band maximum (VBM) and conduction band minimum (CBM) occur at $\sim$0.3$\Gamma$L
and $\sim$0.9$\Gamma$L resulting in two electron and two hole pockets. The band 
structure has an interesting feature near the $\Gamma$ point where the VBM 
has a local minimum and the CBM has a local maximum. These bands
could produce hole and electron pockets if the chemical potential
is moved, for example by doping.

The projected DOS in Figure \ref{fig:bandstructure}(b) clearly shows the 
pseudogap structure with finite DOS at the Fermi level ($E_F$). From the 
projected DOS, one can see that there is strong hybridization between Sb and 
As orbitals. Hybridization of Sb $s$ and As $s$ orbitals gives rise to two 
bands in the DOS from -13.5~eV to -6.5~eV, separated by a gap of $\sim$1~eV. 
The similar picture also applies to the hybridization between Sb $p$ and As $p$ 
orbitals. However, the separation between the bands is small, giving rise to 
an overlap region, forming the pseudogap.

Comparing the band structure of SbAs with those of pure Sb and As 
(Figures \ref{fig:bandstructure}(c-e)) reveals common features over a large 
energy scale, but salient differences appear near the pseudogap region. The splitting of the 
valence and conduction bands at the $L$ point in SbAs is much larger compared to 
the pure compounds, which results in a larger direct
band separation probed by IR reflectivity. The effect of SOI in SbAs is 
stronger than those in pure Sb and As due to the lowering of 
symmetry to non-centrosymmetric $R3m$. Our structural refinements
confirm the presence of anti-site disorder in the SbAs samples which
we measured transport and reflectivity. Further investigation
is required to determine how this disorder affects the
pseudogap structure and whether it can be tuned by annealing
or doping.

Starting from the DFT electronic band structure we calculated the
thermopower $S$. For the nominally 
undoped system we find that the thermopower is positive 22 $\mu$V/K, in
good agreement with our experimental value of 25$\mu$/K. However the calculated
$S$ is nearly $T$-independent in the range 300 K $< T <$ 550 K,
rising from 22 to 26 $\mu$V/K, whereas the experimental value increases from
25 to 40 $\mu$/K. The origin of this strong $T$ dependence is not clear.
The effect of the structural transition around 550 K
on the electronic structure near the Fermi energy and consequent 
change in the chemical potential $\mu$ with temperature and hence $S$ merits further
theoretical study.


\section{Conclusions} 

We find that the compound SbAs forms as an ordered GeTe-type structure
with 80/20 ordering upon cooling from just below the solidus
temperature, as confirmed by
single-crystal x-ray diffraction, high-resolution synchrotron
diffraction, and PDF refinements. The low-$r$ PDF data precludes
the possibility of nanoscale phase separation of Sb and As.
Transport measurements confirm semimetallic behavior analogous
to the end members, with the exception of a direct gap splitting
that is suggested by IR reflectivity to be larger than that of Sb
or As. First-principles calculations indeed find an opening of the 
direct gap around the $L$ point due to subtle changes in
the $p$ orbital hybridization caused by lower symmetry. 

The chemical order-disorder transition is found to be around
550 K by Rietveld refinements to high-temperature synchrotron 
diffraction data. This raises the possibility that
improved ordering may be attained by slow-cooling or annealing below this temperature.
The maximum in thermal conductivity observed by Ohyama
corresponds to ordered SbAs, so the ability to tune this
behavior by control of chemical ordering arises.
Furthermore, a reinvestigation of the Sb$_{1-x}$As$_x$ phase space
is warranted. The related systems Bi$_{1-x}$Sb$_x$ and As$_{1-x}$P$_x$
may also deserve more detailed study. While a wealth of experimental
data finds no miscibility gap in Bi$_{1-x}$Sb$_x$, the behavior of As$_{1-x}$P$_x$ is comparatively
unknown---contradictory reports of the hypothetical compound
AsP have been summarized by Karakaya and Thompson.\cite{karakaya_as-p_1991}

\section{Acknowledgments}
Work at Argonne National Laboratory is supported by UChicago Argonne,
a U.S. DOE Office of Science Laboratory, operated under Contract 
No. DE-AC02-06CH11357.
 This work utilized NPDF at the
Lujan Center at Los Alamos Neutron Science Center, funded by the DOE Office of
Basic Energy Sciences and operated by Los
Alamos National Security LLC under DOE Contract DE-AC52-06NA25396. 

\bibliography{sbas}

\end{document}